\documentclass{JHEP3}

\newcommand{\RR}{{\mathbb R}}

\newcommand{\sgn}{\mbox{sgn}}
\usepackage{latexsym,amsfonts,amsmath,amssymb,graphics,amsthm,eucal,graphicx}
\usepackage{multirow,epsfig,amsmath}

\title{Generalized Superconformal Index for Three Dimensional Field Theories}
\author{Anton Kapustin\\ California Institute of Technology \\ Email: \email{kapustin@theory.caltech.edu}}
\author{Brian Willett\\ California Institute of Technology \\ Email: \email{bwillett@caltech.edu}}

\abstract{We introduce a generalization of the $S^2 \times S^1$ superconformal index where background gauge fields with magnetic flux are coupled to the global symmetries of the theory.  This allows one to gauge a global symmetry at the level of the index, which we use to show the matching of the superconformal index for $\mathcal{N}=2$ SQED with $N_f$ flavors and its mirror dual.}
\preprint{68-2840}

\begin{document}

\section{Introduction}

In this note we discuss a generalized version of the superconformal index on $S^2 \times S^1$ in which the chemical potentials for global $U(1)$ symmetries are supplemented by a new discrete parameter.  One can interpret these chemical potentials as parameterizing the Wilson line along the $S^1$ of a background gauge field which couples to these global symmetries, and then this discrete parameter is the monopole number of the background gauge field configuration.  In principle, computing the index of a theory as a function of these extra parameters should give more information about the theory, and, for example, could provide a  stronger test of dualities.

In field theory, one often gauges a global symmetry of a theory by coupling its conserved currents to a new gauge field.  This can be used, for example, to obtain new dualities from old ones, since if one knows how a symmetry maps across a duality, gauging that symmetry on both sides should give a new pair of equivalent theories.  One advantage of the generalized superconformal index is that this procedure of gauging global symmetries can be performed at the level of the index.  Namely, one simply integrates over the chemical potential and sums over the discrete parameter for the symmetry in question.  One example where this is useful is in deriving mirror symmetry for $\mathcal{N}=2$ SQED with $N_f>1$ flavors from the $N_f=1$ case.  By mirroring the formal proof \cite{Kapustin:1999ha} at the level of the index, we are able to prove the matching of superconformal indices for these theories.  For completeness, we also use a similar procedure to prove the equality of the corresponding $S^3$ partition functions.

\section{Generalized Superconformal Index}

\subsection{Ordinary Index}

Before showing how the superconformal index needs to be modified if one is interested in gauging global symmetries, let us review the ordinary superconformal index.  We will also rewrite it in a more convenient form.

The superconformal index computes the following twisted partition function for a general $\mathcal{N}=2$ superconformal theory on $S^2 \times S^1$ \cite{Kim:2009wb,Imamura:2011su,Krattenthaler:2011da}:

\begin{align} \mbox{tr} \bigg( (-1)^F e^{\beta H } x^{\Delta + j_3} \prod_a {t_a}^{F_a} \bigg) \end{align}
Here $\Delta$ is the energy, $R$ is the $R$-charge, $j_3$ is the third component of the angular momentum rotating $S^2$, the $F_a$ run over the global flavor symmetry generators, and:
\begin{align} H = \{ Q, Q^\dagger \} = \Delta - R - j_3 \end{align}
for a certain supercharge $Q$ in the $\mathcal{N}=2$ superconformal algebra.  By the usual arguments, this implies the index is independent of $\beta$, although it will be a non-trivial function of the chemical potentials $x,t_a$.

In \cite{Imamura:2011su} this index was computed for theories which contain matter of arbitrary superconformal $R$ charge $\Delta$.\footnote{Specifically, it was shown in \cite{Imamura:2011su} that one can treat $\Delta$ as a free parameter by suitably modifying supersymmetry transformations on
$S^2\times S^1$. The price one pays for this is that one looses the
interpretation of the index in terms of counting of superconformal
primaries on $\RR^3$. However, for purposes of testing dualities this is
unimportant.}  One first computes the single particle index:

\begin{align*} \label{spi} \mbox{ind}(e^{i h_j},s_j;t_a;x) = - \sum_{\alpha \in ad(G)} e^{i \alpha(h)} x^{2|\alpha(s)|} + \;\;\;\;\;\;\;\;\;\;\;\;\;\;\;\;\;\;\;\; \end{align*}

\begin{align} + \sum_\Phi \sum_{\rho \in R_\Phi} \bigg( e^{i \rho(h)} \prod_a {t_a}^{f_a(\Phi)} \frac{x^{2 |\rho(s)| + \Delta_\Phi}}{1-x^2} - e^{-i \rho(h)} \prod_a {t_a}^{-f_a(\Phi)} \frac{x^{2 |\rho(s)| + 2-\Delta_\Phi}}{1-x^2} \bigg) \end{align}
Here the first term is the contribution of the vector multiplets, and $\alpha$ runs over the roots of the Lie algebra, while $h$, with components $h_j$, runs over the maximal torus of the group, and parameterizes the $S^1$ Wilson line of the gauge field.  The parameter $s$, with components $s_j$, takes values in the Cartan of the gauge group, and parametrize the GNO charge of the monopole configuration of the gauge field \cite{Kim:2009wb,Borokhov:2003yu}.  In the unitary case that we consider here, the $s_j$ run over half-integers.  

The second term in (\ref{spi}) is the contribution of the chiral multiplets, labeled by $\Phi$.  Here $\rho$ runs over the weights of the representation $R_\Phi$ of the gauge group in which $\Phi$ sits, and $\Delta_\Phi$ is the superconformal $R$-charge of $\Phi$, which, as a consequence of the $\mathcal{N}=2$ superconformal algebra, is also equal to its scaling dimension.  The $t_a$ parametrize the maximal torus of the global symmetry group, and $f_a(\Phi)$ is the charge of $\Phi$ under the $U(1)$ subgroup corresponding to $t_a$.

Next we construct the full index from the single particle index:

\begin{align} \label{fullindex} I(t_a;x) = \sum_{s_j} \frac{1}{Sym} \int e^{-S_{CS}(h,s)} e^{i b_0(h)} x^{\epsilon_0} \prod_a {t_a}^{q_{0a}} \exp \bigg( \sum_{n=1}^\infty \frac{1}{n} \mbox{ind} ( {z_j}^n, s, x^n, {t_a}^n) \bigg) \prod_j \frac{dz_j}{2 \pi i z_j} \end{align}
where $z_j=e^{i h_j}$ each run over the unit circle in the complex plane, and:
\begin{align*} \epsilon_0 &= \sum_\Phi (1- \Delta_\Phi) \sum_{\rho \in R_\Phi} |\rho(s)| - \sum_{\alpha \in ad(G)} |\alpha(s)| \\
q_{0a} &= - \sum_\Phi \sum_{\rho \in R_\Phi} |\rho(s)| f_a(\Phi) \\
b_0(h) &= - \sum_\Phi \sum_{\rho \in R_\Phi} |\rho(s)| \rho(h) \end{align*}
The origin of these factors is explained in \cite{Imamura:2011su}.  Also, ``$Sym$'' is a symmetrization factor arising for nonabelian groups which depends on the magnetic flux.  Specifically, the gauge group is generically broken by monopoles to a subgroup $\otimes_k G_k$, and we define $Sym = \prod_k \mbox{Rank}(G_k)!$.  It can also be written as:

\[ Sym = \prod_{j=1}^{Rank(G)} \bigg( \sum_{k=1}^{Rank(G)} \delta_{s_j,s_k} \bigg) \]
In addition, if a Chern-Simons term is present, it contributes a factor:
\begin{align}\label{cscs} e^{-S_{CS}(h,s)} = e^{-2 i Tr_{CS}(h s)} \end{align}
where $\mbox{Tr}_{CS}$ is the trace containing the Chern-Simons level.  For example, for a $U(N)$ gauge group, a level $k$ Chern-Simons term contributes:

\begin{align} e^{i k \sum_j h_j s_j} = \prod_{j=1}^{N_c} z^{k s_j} \end{align}

We will find it convenient to rewrite the integrand in (\ref{fullindex}) as a product of contributions from the different multiplets.  First, note that the single particle index enters via the so-called plethystic exponential:

\begin{align} \exp \bigg( \sum_{n=1}^\infty \frac{1}{n} \mbox{ind} ( {z_j}^n, s, x^n, {t_a}^n) \bigg) \end{align}
It will be convenient to rewrite this using the $q$-product, defined for $n$ finite or infinite: 

\begin{align} (z;q)_n = \prod_{j=0}^{n-1} (1 - z q^j) \end{align}
Specifically, consider a single chiral field $\Phi$, whose single particle index is given by:
\begin{align} \sum_{\rho \in R_\Phi} \bigg( e^{i \rho(h)} {t_a}^{f_a(\Phi)} \frac{x^{2 |\rho(s)|  + \Delta_\Phi}}{1-x^2} - e^{-i \rho(h)} {t_a}^{-f_a(\Phi)} \frac{x^{2 |\rho(s)| + 2-\Delta_\Phi}}{1-x^2} \bigg) \end{align}
Then we can write the plethystic exponential of this as follows:
\begin{align} \prod_{\rho \in R_\Phi} \exp \bigg( \sum_{n=1}^\infty \frac{1}{n} \bigg( e^{i n\rho(h)} {t_a}^{n f_a(\Phi)} \frac{x^{2 n |\rho(s)|  + n \Delta_\Phi}}{1-x^{2n}} - e^{-i n \rho(h)} {t_a}^{-n f_a(\Phi)} \frac{x^{2 n |\rho(s) | + 2n - n \Delta_\Phi}}{1-x^{2n}} \bigg) \bigg) \end{align}
By rewriting the denominator as a geometric series and interchanging the order of summations, one finds that this becomes:
\begin{align} \prod_{\rho \in R_\Phi} \frac{(e^{-i \rho(h)} {t_a}^{-f_a(\Phi)} x^{2 |\rho(s)| + 2 -\Delta_\Phi} ; x^2)_\infty}{(e^{i \rho(h)} {t_a}^{f_a(\Phi)} x^{2 |\rho(s)| + \Delta_\Phi} ; x^2)_\infty} \end{align}

The full index will involve a product of such factors over all the chiral fields in the theory, as well as the contribution from the gauge multiplet.  It is given by:
\begin{align}\label{2.11} I(t_a;x) = \sum_{s_j} \frac{1}{Sym} \int e^{-S_{CS}(h,s)} Z_{gauge}(z_j,s_j;x) \prod_\Phi Z_\Phi(z_j,s_j; t_a;x) \prod_j \frac{dz_j}{2 \pi i z_j}\end{align}
where:

\begin{align*} Z_{gauge}(z_j,s_j;x) = \prod_{\alpha \in ad(G)} x^{-|\alpha(s)| } \bigg(1 - e^{i \alpha(h)} x^{2 |\alpha(s)|} \bigg)  \end{align*}

\begin{align*} Z_\Phi(z_j=e^{i h_j},s_j; t_a;x) = \prod_{\rho \in R_\Phi}  \bigg(x^{(1- \Delta_\Phi) } \prod_j e^{-i \rho(h)} \prod_a {t_a}^{-f_a(\Phi)} \bigg)^{|\rho(s)|} \frac{(e^{-i \rho(h)} {t_a}^{-f_a(\Phi)} x^{2 |\rho(s)| + 2 -\Delta_\Phi} ; x^2)_\infty}{(e^{i \rho(h)} {t_a}^{f_a(\Phi)} x^{2 |\rho(s)| + \Delta_\Phi} ; x^2)_\infty} \end{align*}
In particular, when the gauge group is abelian, as in the cases we will discuss below, its contribution is trivial.

\subsection{Generalized Index}

Note that $z_j$ and $t_a$ appear on a very similar footing in (\ref{2.11}).  The reason for this is that, in similarity to the $z_j$, we can think of the chemical potentials $t_a$ as parametrizing the $S^1$ Wilson lines for fixed, flat background gauge fields which couple to the global symmetries of the theory.  Since they are fixed, we do not integrate over the $t_a$ as we do for the $z_j$, and since they are flat, there is not a corresponding magnetic flux, analagous to $s_j$, for these background gauge fields.

Now suppose we gauge one of these global $U(1)$ symmetries $F_a$.  That is, we introduce new dynamical gauge fields which couple to its conserved flavor current.  It is easy to see that the new integration variable $z_a$ is introduced by making the replacement $t_a \rightarrow z_a$.  In addition, one must introduce a new variable $s_a$ corresponding to its magnetic flux.  We see that the appropriate change here is to make the following replacement in (\ref{2.11}):

\begin{align} \rho(s) \rightarrow \rho(s) + f_a(\Phi) s_a \end{align}
One now integrates over $z_a$ and sums over the new discrete parameter $s_a$.

But now the definition of the generalized superconformal index is clear.  One simply makes the replacement above for all flavor symmetries, but does not integrate and sum over the corresponding parameters.  Rather, they are the variables on which this generalized index depends. 

If one wants to gauge a flavor symmetry, one simply integrates and sums over the corresponding parameters, introducing the appropriate contribution for the new vector multiplet if it is nonabelian.  Note the sum over the discrete parameter corresponding to a gauge group is a necessary ingredient in the superconformal index, which is why this generalized index is necessary if one intends to gauge global symmetries.

Let us continue to call the continuous parameter for the flavor symmetries $t_a$, and denote the new discrete parameter by $n_a$ to distinguish it from the $s_j$ which are summed over.  Then, to summarize, we can write the generalized superconformal index as follows:

\begin{align} I(t_a,n_a;x) = \sum_s \frac{1}{Sym} \int e^{-S_{CS}(h,s)} Z_{gauge}(z_j,s_j;x) \prod_\Phi Z_\Phi(z_j,s_j; t_a;x) \prod_j \frac{dz_j}{2 \pi i z_j}\end{align}
where the gauge contribution is unchanged, and the contribution of a chiral multiplet becomes:

\begin{align*} Z_\Phi(z_j=e^{i h_j},s_j; t_a,n_a;x) = \prod_{\rho \in R_\Phi}  \bigg(x^{(1- \Delta_\Phi) } \prod_j e^{-i \rho(h)} \prod_a {t_a}^{-f_a(\Phi)} \bigg)^{|\rho(s) + \sum_a f_a(\Phi) n_a|} \times \\
\times \frac{(e^{-i \rho(h)} {t_a}^{-f_a(\Phi)} x^{2 |\rho(s)+ \sum_a f_a(\Phi) n_a| + 2 -\Delta_\Phi} ; x^2)_\infty}{(e^{i \rho(h)} {t_a}^{f_a(\Phi)} x^{2 |\rho(s)+ \sum_a f_a(\Phi) n_a| + \Delta_\Phi} ; x^2)_\infty} \end{align*}
As with the $s_j$, the $n_a$ should take values in $\frac{1}{2}\mathbb{Z}$.

There is another type of global symmetry we can consider.  For every $U(N)$ factor in the gauge group, there is a conserved topological current $J=\star \mbox{Tr} F$ corresponding to a global symmetry, which we call $U(1)_J$.  To determine the contribution of this symmetry to the generalized index, we make an analogy with the procedure we used above for flavor symmetries and couple the $U(1)_J$ symmetry to a background vector multiplet $V_{BG}$.  Coupling $J$ to a background vector multiplet is equivalent to including an $\mathcal{N}=2$ $BF$ term \cite{Kapustin:1999ha,Brooks:1994nn}:

\begin{align} \frac{1}{2\pi} \int d^3x d^2 \theta \mbox{Tr}(\Sigma) V_{BG} = \int d^3 x (A_{BG} \wedge \mbox{Tr} dA + ... ) \end{align}
This can be thought of as an off-diagonal Chern-Simons term, and we can read off the contribution of such a term from (\ref{cscs}).  For example, for a $U(1)$ gauge group with parameters $(z,s)$, one finds an extra factor in the integrand of the index of:

\begin{align} z^{2n} w^{2s} \end{align}
where $w$ and $n$ are respectively the continuous and discrete parameters for the background $U(1)$ gauge field.

Before moving on, we make the following observation, which also applies to the ordinary superconformal index. If we make the replacement $t_a \rightarrow t_a x^{c_a}$, the matter contribution to the superconformal index becomes:

\begin{align*} Z_\Phi(z_j,s_j; t_a;x) = \prod_{\rho \in R_\Phi}  \bigg(x^{(1- \Delta_\Phi - \sum_a c_a f_a(\Phi)) } \prod_j e^{-i \rho(h)} \prod_a {t_a}^{-f_a(\Phi)} \bigg)^{|\rho(s)|} \times \\ \times \frac{(e^{-i \rho(h)} {t_a}^{-f_a(\Phi)} x^{2 |\rho(s)| + 2 -\Delta_\Phi - \sum_a c_a f_a(\Phi)} ; x^2)_\infty}{(e^{i \rho(h)} {t_a}^{f_a(\Phi)} x^{2 |\rho(s)| + \Delta_\Phi + \sum_a c_a f_a(\Phi)} ; x^2)_\infty} \end{align*}
Note that the dimension $\Delta_\Phi$ always appears in the combination $\Delta_\Phi + \sum_a c_a f_a(\Phi)$.  This means that if one shifts the $R$-symmetry by a combination $\sum_a c_a Q_a$ of the flavor symmetries, the appropriate modification of the superconformal index is to take $t_a \rightarrow t_a x^{c_a}$.  A similar phenomenon was observed with the $S^3$ partition in \cite{Jafferis:2010un}.  In other words, although $\Delta_\Phi$ is a free parameter, a shift in it can be absorbed into a redefinition of the $t_a$.  Thus we may as well choose $\Delta_\Phi$ according to the UV dimensions of the fields.

\section{Abelian Mirror Symmetry}

In this section we consider $\mathcal{N}=2$ abelian mirror symmetry, and demonstrate the matching of both the superconformal index and the $S^3$ partition function for mirror pairs.  As shown in \cite{Kapustin:1999ha}, one can formally prove mirror symmetry for SQED with a general number of flavors by starting with the $N_f=1$ case and gauging global symmetries.  As we will see below, one can perform an analogous argument to prove the equality of the $S^3$ and $S^2 \times S^1$ partition functions.

\subsection{The Formal Argument}

Let us review how the formal argument works.  For a single flavor, mirror symmetry states that the superconformal IR fixed point of the following theories are equivalent:

\begin{itemize}
\item $\mathcal{N}=2$ SQED with a single flavor.  Here a flavor consists of two chiral fields, $Q$ and $\tilde{Q}$, with charges $1$ and $-1$ respectively.  This theory has an axial $U(1)_A$ flavor symmetry which rotates $Q$ and $\tilde{Q}$ by the same phases\footnote{The vector symmetry, which would rotate them by opposite phases, is gauged here.} as well as a topological $U(1)_J$ symmetry.
\item The $XYZ$ theory, a theory of three chiral fields $X,Y$, and $Z$ interacting via the superpotential $W=XYZ$.  We will find it convenient to rename the three fields to $q,\tilde{q},$ and $S$.  There is a $U(1)^2$ symmetry, which we will parameterize as $U(1)_V \times U(1)_A$, where the fields are charged as:

\begin{align} q \rightarrow (1,1) , \;\;\;\; \tilde{q} \rightarrow (-1,1), \;\;\;\; S \rightarrow (0,-2) \end{align} 
\end{itemize}

Mirror symmetry tells us these theories have equivalent IR fixed points, and moreover, that the symmetries of the two theories are identified via (here the LHS denotes symmetries of the first theory):

\begin{align} U(1)_J \leftrightarrow U(1)_V \\ \nonumber
U(1)_A \leftrightarrow {U(1)_A} \end{align}
More precisely, the $U(1)_A$ current is mapped with a change of sign \cite{Tong:2000ky}.  In addition, the chiral operator $Q \tilde{Q}$ in the first theory is identified with $S$ in the dual.

One can obtain the $\mathcal{N}=4$ version of SQED by adding a new adjoint (uncharged) chiral $\tilde{S}$ which couples to $Q \tilde{Q}$.  On the dual side, this corresponds to adding a superpotential $S \tilde{S}$ which makes these two fields massive and so we can integrate them out, leaving the theory of the free (twisted) hypermultiplet formed from $q$ and $\tilde{q}$.

In order to obtain the duality for a general number of flavors, we will use the following fact.  Consider a theory with an abelian gauge field $A$, and corresponding topological current $J=\star dA$.  To gauge this current, we introduce a new gauge field $B$ which couples via a $BF$ term:

\begin{align} \label{BF} B \wedge dA \end{align}
Then, as discussed in \cite{Witten:2003ya}, integrating over $B$ imposes that $F=0$ on an arbitrary manifold.  This holds at the level of supermultiplets as well, namely, if we couple the current to an entire background $\mathcal{N}=2$ vector multiplet via the supersymmetric extension of the BF term \cite{Kapustin:1999ha,Brooks:1994nn}, then one finds that the original vector multiplet is set to zero.  More generally, for every $U(N)$ factor in the gauge group, there is a $U(1)_J$ symmetry with current $\star \mbox{Tr} F$, and gauging this symmetry sets the trace part of the gauge multiplet to zero, thereby reducing this factor to its $SU(N)$ subgroup.

We obtain the duality for general $N_f$ as follows.  We can construct $\mathcal{N}=2$ SQED with $N_f$ flavors by starting with $N_f$ free hypermultiplets and gauging the sum of the $U(1)_V$ currents.  To get the dual, we use the $\mathcal{N}=4$ duality to argue that this is the same as starting with $N_f$ copies of SQED with an adjoint chiral, and gauging the sum of the $U(1)_J$ currents.  From above, we see this ungauges the diagonal $U(1)$ of the gauge group, and so we are left with a theory whose gauge group is $U(1)^{N_f-1}$, namely, it is the kernel of the homomorphism from $U(1)^{N_f}$ to $U(1)$ which multiplies the $N_f$ elements together.  In addition, there are $N_f$ flavors $(Q_a,\tilde{Q}_a)$ charged as $(1,0,0,...,0),(0,1,0,...,0),(0,0,0,...,1)$.  Finally, there are $N_f$ uncharged chirals $S_a$ which couple via the superpotential $\sum_a Q_a \tilde{Q}_a S_a$

Actually, this description of the latter theory is not the usual one for the mirror of SQED.  To correct this, we define new gauge fields $\hat{A}_a$ implicitly by $A_a = \hat{A}_a - \hat{A}_{a+1}$.  This is possible because the $A_a$ sum to zero, but it only determines the $\hat{A}_a$ up to an $a$-independent shift, which decouples from the theory.  Thus we get a theory with gauge group $U(1)^{N_f}/U(1)^{diag}$ and matter charged as $(1,-1,0,...,0)$, $(0,1,-1,...,0),$ $(-1,0,0,...,1)$, which is the usual presentation of the mirror of $\mathcal{N}=2$ SQED with $N_f$ flavors.

We can also derive how the currents for the global symmetries map in the general $N_f$ duality from the mapping in the $N_f=1$ case.  The $a$th flavor on the SQED side has a $U(1)_V$ and a $U(1)_A$ current.  The former is mapped to the topological current $J_{T,a} = \star F_a$ (or, after the redefinition of the previous paragraph, to the difference $\hat{J}_{T,a} - \hat{J}_{T,a+1}$), while the latter is mapped, with a change of sign, to the $U(1)_A$ current acting on the chirals $(Q_a,\tilde{Q}_a,\tilde{S}_a)$.  Finally, in SQED, the diagonal sum of the $U(1)_V$ currents has been gauged, and this gauge field has a $U(1)_J$ symmetry.  On the dual side, the $U(1)_J$ for the diagonal sum of the $U(1)$ gauge currents has been gauged, which corresponds to ungauging this sum, and promoting the symmetry rotating all $N_f$ flavors by the same phase to a global symmetry.  Then, from \cite{Witten:2003ya}\footnote{Specifically, this follows from the fact that, in the notation of that paper, $S^2=-1$.}, one sees that the $U(1)_J$ current in SQED is identified with this flavor current in the dual, up to a change of sign.  A more careful analysis shows there is also a numerical factor, and $J_T \leftrightarrow -\frac{1}{N_f} J_V$.  Note that these global symmetries are enhanced in SQED to the non-abelian group $SU(N_f)_V \times U(N_f)_A \times U(1)_J$, but only the maximal torus of this is visible in the UV description of the dual theory.

\subsection{$S^3$ Partition Function}

Before showing how this procedure works at the level of the superconformal index, it will be instructive to carry it out with the somewhat simpler $S^3$ partition function.  In both cases, the arguments will be structurally identical the formal arguments of the previous section.  However, in these cases one can rigorously establish the $N_f=1$ base case by other means \cite{Jafferis:2010un,Hama:2010av}.  The matching of $S^3$ partition functions in the $\mathcal{N}=4$ case was shown by a similar method in \cite{Kapustin:2010xq}.

The $S^3$ partition function localizes to an integral over the Cartan of the gauge group of the theory.  For example, a $U(N)$ gauge group contributes $N$ integration variables $\lambda_i$, $i=1,...,N$, corresponding to the eigenvalues of the adjoint scalar in the vector multiplet, and the integration measure is:

\begin{align} \int d^N \lambda \prod_{i<j} (2 \sinh \pi(\lambda_i - \lambda_j))^2 \end{align}
For abelian gauge groups, such as the ones we will consider here, this measure is simply $d \lambda$.  Meanwhile, a chiral multiplet of dimension $\Delta$ in a representation $R$ of the gauge group contributes a factor of:

\begin{align} \prod_{\rho \in R} e^{\ell(1 - \Delta + i \rho(\sigma))} \end{align}
where $\sigma = \mbox{diag}(\lambda_1,...,\lambda_N)$ and $\rho$ runs over the weights of the representation.  The function $\ell(z)$ is given by \cite{Jafferis:2010un,Hama:2010av}:

\begin{align}\label{ell} \ell(z) = - z \log ( 1 - e^{2 \pi i z} ) + \frac{i}{2} ( \pi z^2 + \frac{1}{\pi} \mbox{Li}_2(e^{2 \pi i z})) - \frac{i \pi}{12} \end{align}

In addition to the gauge symmetries, this matter couples to global flavor currents.  Much as with the generalized superconformal index, one can consider coupling these currents to background gauge fields.  Then the scalar eigenvalues for these background gauge fields are interpreted as real mass parameters corresponding to the global symmetries.  As discussed in \cite{Jafferis:2010un}, one can pick $\Delta$ according to the UV dimensions of the fields, and more general dimensions can be obtained by allowing these mass parameters to become complex.

In addition to mass parameters, one can include FI terms by gauging the topological $U(1)_J$ symmetry.  This enters the matrix model as a factor in the integrand of the form:

\begin{align} e^{2 \pi i \eta \sum_j \lambda_j} \end{align}
Note that integrating over $\eta$ introduces a delta function constraint.  This reflects what we found in the previous section, where this operation imposed $\mbox{tr} F=0$.

Let us investigate the $S^3$ partition function for the theories involved in abelian mirror symmetry.  We will first establish the $N_f=1$ case, and then derive the general case by mirroring the arguments of the previous section at the level of the matrix model.

The matching of the partition functions for the $N_f=1$ duality was shown in \cite{Jafferis:2010un,Hama:2010av}.  It also follows from the results of \cite{Willett:2011gp}.  Namely, the case $N_c=N_f=1$ of the Aharony-Seiberg duality considered in that paper is precisely the $N_f=1$ case of mirror symmetry.  The fields $V_+,V_-,$ and $M$ of the former are identified with $q,\tilde{q},S$ in the latter.

Using the results of these papers, we find that the partition function of $N_f=1$ SQED, deformed by an axial mass $\mu$ and FI term $\eta$, which is given by:

\begin{align} Z_{N_f=1}(\eta,\mu) = \int d \lambda e^{2 \pi i \eta \lambda} e^{\ell(1/2 + i \lambda + i \mu) - \ell(1/2 - i \lambda + i \mu)} \end{align}
and the partition function of the XYZ theory, deformed by vector and axial masses $\tilde{m}$ and $\tilde{\mu}$, given by:

\begin{align} Z_{XYZ}(\tilde{m},\tilde{\mu}) = e^{\ell(-2 i \tilde{\mu})}e^{\ell(1/2 + i \tilde{m} + i \tilde{\mu}) + \ell(1/2 - i \tilde{m} + i \tilde{\mu}) } \end{align}
are related as expected by the mapping of symmetries, namely:

\begin{align} Z_{N_f=1}(\eta,\mu) = Z_{XYZ}(\eta,-\mu) \end{align}
or, explicitly:
\begin{align} \label{expdua}\int d \lambda e^{2 \pi i \eta \lambda} e^{\ell(\frac{1}{2} + i \lambda + i \mu) + \ell(\frac{1}{2} - i \lambda + i \mu)} = e^{\ell(2 i \mu)} e^{\ell(\frac{1}{2} + i \eta - i \mu)+ \ell(\frac{1}{2} - i \eta - i \mu)} \end{align} 

As described above, we can eliminate $S$ by coupling it to a new chiral $\tilde{S}$ by a superpotential $S \tilde{S}$, and we obtain the duality between a free twisted hypermultiplet and $\mathcal{N}=4$ SQED.  Note that the superpotential $S \tilde{S}$ must be flavor neutral and have dimension $2$, which means $S$ and $\tilde{S}$ must come in conjugate representations of the flavor group and have $\Delta_S + \Delta_{\tilde{S}}=2$.  Since this superpotential implies the fields are absent from the IR theory, we must have:

\[ Z_\Delta(m_a) Z_{2-\Delta}(-m_a) = 1 \]
Inspecting (\ref{2.11}) and (\ref{ell}), we see this is true for both the $S^2 \times S^1$ index and the $S^3$ partition function.  Thus this operation corresponds in the matrix model to simply moving the factor corresponding to $S$ from one side of (\ref{expdua}) to the other.  We obtain the identity:

\begin{align}\label{usdual} e^{\ell(-2 i \mu)}\int d \lambda e^{2 \pi i \eta \lambda} e^{\ell(\frac{1}{2} + i \lambda + i \mu) + \ell(\frac{1}{2} - i \lambda + i \mu)} =  e^{\ell(\frac{1}{2} + i \eta - i \mu)+ \ell(\frac{1}{2} - i \eta - i \mu)} \end{align}
where the LHS is the partition function of $\mathcal{N}=4$ SQED, and the RHS that of a free twisted hypermultiplet, each deformed by an axial mass term.  The axial masses must be set to zero to preserve $\mathcal{N}=4$ supersymmetry, although we will find this more general identity useful.

Now we obtain the duality for general $N_f$ as follows.  We start with $N_f$ copies of a free hypermultiplet, deformed by axial mass terms:

\begin{align}\label{nfduaa} \prod_{a=1}^{N_f} e^{\ell(\frac{1}{2} + i \tilde{m}_a + i \mu_a)+ \ell(\frac{1}{2} - i \tilde{m}_a + i \mu_a)} \end{align}
Then we gauge the sum of the $U(1)_V$ symmetries for all the hypermultiplets.  Redefining parameters by $\tilde{m}_a = m_a + \lambda$, where $\sum_a m_a = 0$, and introducing an FI term $\eta$ for the gauge group, this gives:

\[ \int d\lambda e^{2 \pi i \eta \lambda}  \prod_{a=1}^{N_f} e^{\ell(\frac{1}{2} + i \lambda + i m_a + i \mu_a)+ \ell(\frac{1}{2} - i \lambda - i m_a + i \mu)} \]
This is the partition function for $\mathcal{N}=2$ SQED with $N_f$ flavors.

To get the dual, we use (\ref{usdual}) to rewrite (\ref{nfduaa}) as:

\[ \prod_{a=1}^{N_f} e^{\ell(2 i \mu_a)}\int d \lambda_a e^{2 \pi i \tilde{m}_a \lambda_a} e^{\ell(\frac{1}{2} + i \lambda_a - i \mu_a) + \ell(\frac{1}{2} - i \lambda_a - i \mu_a)} \]
Redefining parameters as above and gauging the corresponding symmetry, we obtain:

\[ \int d \lambda e^{2 \pi i \eta \lambda} \prod_{a=1}^{N_f} e^{\ell(2 i \mu_a)}\int d \lambda_a e^{2 \pi i (\lambda +m_a) \lambda_a} e^{\ell(\frac{1}{2} + i \lambda_a - i \mu_a) + \ell(\frac{1}{2} - i \lambda_a - i \mu_a)} \]

\[ = \int \delta( \sum_a \lambda_a + \eta) \prod_{a=1}^{N_f} e^{\ell(2 i \mu_a)}\int d \lambda_a e^{2 \pi i m_a \lambda_a} e^{\ell(\frac{1}{2} + i \lambda_a - i \mu_a) + \ell(\frac{1}{2} - i \lambda_a - i \mu_a)} \]

\begin{align}\label{delt} = \int \delta( \sum_a \lambda_a) \prod_{a=1}^{N_f} e^{\ell(2 i \mu_a)}\int d \lambda_a e^{2 \pi i m_a \lambda_a} e^{\ell(\frac{1}{2} + i \lambda_a - i \frac{\eta}{N_f} - i \mu_a) + \ell(\frac{1}{2} - i \lambda_a + i \frac{\eta}{N_f}- i \mu_a)} \end{align}
where in the last line we have used $\sum_a m_a = 0$.

The final step is to redefine variables to get the standard presentation of the dual theory.  It is straightforward to verify the following relation between measures:

\begin{align}\label{redeff}
\delta(\sum_{a=1}^{N_f} \lambda_a) \prod_{a=1}^{N_f} d \lambda_a = \delta(\frac{1}{N_f}\sum_{a=1}^{N_f} \hat{\lambda}_a) \prod_{a=1}^{N_f} d \hat{\lambda}_a
\end{align}
where the variables are related by $\lambda_a=\hat{\lambda}_a-\hat{\lambda}_{a+1}$.  Applying this identity to (\ref{delt}), we get:

\begin{align}\label{delt2} \int \delta( \frac{1}{N_f} \sum_a \hat{\lambda}_a ) \prod_{a=1}^{N_f} e^{\ell(2 i \mu_a)}\int d \lambda_a e^{2 \pi i (m_a - m_{a-1}) \hat{\lambda}_a} e^{\ell(\frac{1}{2} + i (\hat{\lambda}_a - \hat{\lambda}_{a+1}) - i \frac{\eta}{N_f} - i \mu_a) + \ell(\frac{1}{2} - i (\hat{\lambda}_a - \hat{\lambda}_{a+1}) + i \frac{\eta}{N_f} - i \mu_a)} \end{align}
This is the partition function for the dual theory.  We see that the deformations map as expected, with vector mass terms exchanged with FI terms and axial mass terms mapped to themselves, up to a sign.  This proves the duality at the level of the $S^3$ partition function for arbitrary $N_f$.

\subsection{$S^2 \times S^1$ Superconformal Index}

Let us now consider the (generalized) superconformal index on $S^2 \times S^1$.  As before, we start with the case of $\mathcal{N}=2$ SQED with one flavor.  If we denote the parameters for the $U(1)_A$ and $U(1)_J$ symmetries by $(\alpha,m)$ and $(w,n)$, then using (\ref{2.11}), one finds the generalized superconformal index is given by:

\begin{align} I_{N_f=1}(\alpha,m;w,n;x) = \sum_{s \in \mathbb{Z}/2} \int \frac{dz}{2 \pi i z} z^{2n} w^{2s} (x^{1/2} z^{\pm 1} \alpha^{-1})^{|s \mp m|} \frac{(z^{\pm 1} \alpha^{-1} x^{2|s \mp m|+3/2};x^2)_\infty}{(z^{\pm 1} \alpha x^{2|s \pm m|+1/2};x^2)_\infty} \end{align}
where we use the notation $(z^\pm;q)_n = (z^+;q)_n(z^-;q)_n$.  We also define $(z_1,z_2,...;q)_n =(z_1;q)_n(z_2;q)_n...$.

It was shown in \cite{Krattenthaler:2011da} that the ordinary index, obtained by setting $m=n=0$, agrees with the index of the $XYZ$ model.  However, in order to carry out the procedure described above, one must be able to gauge global symmetries, and so we will need the stronger result that the generalized superconformal indices match.  Actually, if one only wants to show the matching of the ordinary superconformal indices of the general $N_f$ theories and their mirrors, one only needs to let $n$ be non-zero, since we will only gauge the $U(1)_J$ symmetry.  We were unable to prove the matching for non-zero $m$, so we will set $m$ to be zero here.  In the appendix we check the agreement for arbitrary $m$ and $n$ and at the lowest order in the parameter $x$.

We will evaluate this integral by modifying the method of \cite{Krattenthaler:2011da}.  Let us first redefine variables by:

\begin{align} q=x^2, \;\;\; k = 2s, \;\;\; \ell = 2 n , \;\;\; a = \alpha^{-2} q^{1/2}\end{align}
Then the integral we wish to evaluate is:

\begin{align} \sum_{k \in \mathbb{Z}} a^{|k|/2} w^k \int \frac{z^{\ell} dz}{2 \pi i z} \frac{(z^{\pm 1} a^{1/2} q^{|k|/2+1/2};q)_\infty}{(z^{\pm 1} a^{-1/2} q^{|k|/2+1/2};q)_\infty} \end{align}
The poles in the integrand of occur at:

\begin{align} z = (a^{-1/2} q^{|k|/2+j+1/2})^{\pm 1} \end{align}
for non-negative integers $j$.  Now, we must take $|q|<1$ for the infinite $q$-product to make sense, and let us also make the assumption $|q a^{-1} |<1$, so that the unit circle splits the two sets of poles.  Once we have the final answer, we can relax this assumption by analytic continuation.

Because of the $z^\ell$ factor in the integrand, it is convenient to take the poles lying inside and outside the unit circle for $\ell>0$ and $\ell<0$ respectively.  From the assumption above, this corresponds to taking the poles at:

\begin{align} z = (a^{-1/2} q^{|k|/2+j+1/2})^{\sgn(\ell)} \end{align}
and the sum over residues gives:

\begin{align} I_{N_f=1} = \sum_{k \in \mathbb{Z}} \sum_{j=0}^\infty a^{|k|/2} w^k (a^{-1/2} q^{|k|/2+j+1/2})^{|\ell|} \frac{(q^{|k|+j+1},a q^{-j};q)_\infty}{(a^{-1}q^{|k|+j+1},q;q)_\infty (q^{-j};q)_j}\end{align}
It is straightforward to modify the argument in \cite{Krattenthaler:2011da} to show that we may replace all $|k|$'s with $k$'s.  Then, after some simplification, this becomes:

\begin{align} \sum_{j=0}^\infty q^{j |\ell| + |\ell|/2} a^{j-|\ell|/2} \frac{(a;q)_\infty {(qa^{-1};q)_j}^2}{(q;q)_\infty {(q;q)_j}^2} \;_1 \psi_1 \left[ \begin{array}{c} q^{1+j} a^{-1} \\ q^{1+j} \end{array}; q, q^{|\ell|/2} a^{1/2} w \right] \end{align}
where:

\begin{align} \;_1 \psi_1 \left[ \begin{array}{c} A \\ B \end{array}; q, Z \right] = \sum_{m \in \mathbb{Z}} \frac{(A;q)_m}{(B;q)_m} Z^m = \frac{(q,B/A,AZ,q/AZ;q)_\infty}{(B,q/A,Z,B/AZ;q)_\infty} \end{align}
where the second equality is Ramanujan's summation \cite{Krattenthaler:2011da}.  This leaves:

\begin{align} I_{N_f=1} = \sum_{j=0}^\infty q^{j |\ell| + |\ell|/2} a^{j-|\ell|/2}  \frac{(a;q)_\infty {(qa^{-1};q)_j}^2}{(q;q)_\infty {(q;q)_j}^2} \frac{(q,a,q^{1+j+|\ell|/2} a^{-1/2} w, q^{-j-|\ell|/2} a^{1/2} w^{-1};q)_\infty}{(q^{1+j},q^{-j} a, q^{|\ell|/2} a^{1/2} w ,q^{-|\ell|/2} a^{1/2} w^{-1};q)_\infty} \end{align}

\begin{align} = q^{|\ell|/2} a^{-|\ell|/2} \frac{(a,q^{1+|\ell|/2} a^{-1/2} w;q)_\infty}{(qa^{-1},q^{|\ell|/2} a^{1/2} w;q)_\infty} \;_1 \phi_0 \left[ \begin{array}{c} q a^{-1} \\ - \end{array} ; q, q^{|\ell|/2} a^{1/2} w^{-1} \right] \end{align}
where:

\begin{align} \;_1 \phi_0 \left[ \begin{array}{c} A \\ - \end{array} ; q, Z \right] = \sum_{m=0}^\infty \frac{(A;q)_m}{(q,q)_m} Z^m = \frac{(AZ;q)_\infty}{(Z;q)_\infty} \end{align}
using the $q$-binomial theorem.  Plugging this in and simplifying, we arrive at:

\begin{align} I_{N_f=1} = q^{|\ell|/2} a^{-|\ell|/2} \frac{(a;q)_\infty}{(q a^{-1};q)_\infty} \frac{(q^{1+|\ell|/2} a^{-1/2} w^{\pm 1};q)_\infty}{(q^{|\ell|/2} a^{1/2} w^{\pm 1};q)_\infty} \end{align}
Or, in terms of the original variables:

\begin{equation}
\label{ft}
I_{N_f=1}(\alpha,0;w,n;x) = x^{|n|} \alpha^{2 |n|} \frac{(\alpha^{-2} x; x^2)_\infty}{(\alpha^2 x;x^2)_\infty} \frac{(\alpha x^{2|n| + 3/2} w^{\pm 1};x^2)_\infty}{(\alpha^{-1} x^{2|n|+1/2} w^{\pm 1};x^2)_\infty}
\end{equation}

Now consider the XYZ theory.  Here we should take $q,\tilde{q}$, and $S$ to have UV $R$-charges $1/2,1/2$, and $1$ respectively.  The global symmetries are $U(1)_A$ and $U(1)_V$, with corresponding parameters $(\tilde{\alpha},\tilde{m})$ and $(\tilde{\beta},\tilde{n})$.  Then the generalized superconformal index is given by:

\begin{align} I_{XYZ}(\tilde{\alpha},\tilde{m};\tilde{\beta},\tilde{n};x) = (x^{1/2} \tilde{\alpha}^{-1} \tilde{\beta}^{\pm 1})^{|\tilde{m} \mp \tilde{n}|} (\tilde{\alpha}^2)^{2|\tilde{m}|} \frac{(\tilde{\alpha}^{-1} \tilde{\beta}^{\pm 1} x^{2 |\tilde{m} \mp \tilde{n}|+3/2};x^2)_\infty}{(\tilde{\alpha} \tilde{\beta}^{\pm 1} x^{2 |\tilde{m} \pm \tilde{n}|+1/2};x^2)_\infty} \frac{(\tilde{\alpha}^2 x^{4|\tilde{m}|+1};x^2)_\infty}{(\tilde{\alpha}^{-2} x^{4|\tilde{m}|+1};x^2)_\infty} \end{align}

We see that if we set $\tilde{m}=0$ and identify the parameters as dictated by the mapping of symmetries, namely:

\begin{align}\label{1} \tilde{\alpha} = \alpha^{-1}, \;\;\; \tilde{\beta} = w, \;\;\; \tilde{n} = n ,\end{align}
then the expressions (\ref{ft}) and (\ref{1}) match, as expected.

We will now assume the indices match for arbitrary $m$, although we have only shown this in the case $m=0$.  It will be clear from our argument that the special case $m=0$ implies mirror symmetry for the ordinary superconformal index for arbitrary $N_f$.

As before, we can obtain the $\mathcal{N}=4$ version of the duality (with axial deformations) by moving the contribution of $S$ from one side to the other.  The corresponding identity is:

\begin{align}\label{scinf}  (\alpha^{2})^{2|m|}  \frac{(\alpha^{2} x^{4|m|+1};x^2)_\infty}{(\alpha^{-2} x^{4|m|+1};x^2)_\infty} 
 \sum_{s \in \mathbb{Z}/2} \int \frac{dz}{2 \pi i z} z^{2n} w^{2s} (x^{1/2} z^{\pm 1} \alpha^{-1})^{|s \mp m|} \frac{(z^{\pm 1} \alpha^{-1} x^{2|s \mp m|+3/2};x^2)_\infty}{(z^{\pm 1} \alpha x^{2|s \pm m|+1/2};x^2)_\infty} = \end{align}

\begin{align*} = (x^{1/2} \alpha w^{\pm 1})^{|m \pm n|} \frac{(\alpha w^{\pm 1} x^{2 |m \pm n|+3/2};x^2)_\infty}{(\alpha^{-1} w^{\pm 1} x^{2 |m \mp n|+1/2};x^2)_\infty}\end{align*}

Now to get the duality for general $N_f$, we start as before with the generalized superconformal index for $N_f$ free hypermultiplets:

\begin{align}\label{last} \prod_{a=1}^{N_f} (x^{1/2} \alpha_a {w_a}^{\pm 1})^{|m_a \pm n_a|} \frac{(\alpha_a {w_a}^{\pm 1} x^{2 |m_a \pm n_a|+3/2};x^2)_\infty}{({\alpha_a}^{-1} {w_a}^{\pm 1} x^{2 |m_a \mp n_a|+1/2};x^2)_\infty} \end{align}
Now we redefine parameters by $w_a = u_a z$, where $\prod_a u_a =1$, and $n_a = v_a + s$, where $\sum_a v_a =0$ (here $u_a,v_a$ will correspond to the continuous and discrete parameter for the $a$th $U(1)_J$ symmetry) and gauge the sum of the $U(1)_V$ currents, with parameters $w,n$ for the $U(1)_J$ symmetry of the new gauge field:

\[ \sum_{s \in \mathbb{Z}/2} \int \frac{dz}{2 \pi i z} z^{2n} w^{2 s} \prod_{a=1}^{N_f} (x^{1/2} \alpha_a (z u_a)^{\pm 1})^{|n_a + s \pm m_a|} \frac{(\alpha_a (z u_a)^{\pm 1} x^{2 |n_a + s \pm m_a|+3/2};x^2)_\infty}{({\alpha_a}^{-1} (z u_a)^{\pm 1} x^{2 |n_a + s \mp m_a|+1/2};x^2)_\infty} \]
This is the generalized superconformal index for $\mathcal{N}=2$ SQED with $N_f$ flavors.

To get the generalized superconformal index of the dual theory, we use (\ref{scinf}) to rewrite (\ref{last}) as:

\begin{align} \prod_{a=1}^{N_f} (x^{1/2}  {\alpha_a}^{2})^{2|m_a|}  \frac{({\alpha_a}^{2} x^{4|m_a|+1};x^2)_\infty}{({\alpha_a}^{-2} x^{4|m_a|+1};x^2)_\infty} \times \end{align}

\begin{align*}\times \sum_{s_a \in \mathbb{Z}/2} \int \frac{dz_a}{2 \pi i z_a} {z_a}^{2n_a} {w_a}^{2s_a} (x^{1/2} {z_a}^{\pm 1} {\alpha_a}^{-1})^{|s_a \mp m_a|} \frac{({z_a}^{\pm 1} {\alpha_a}^{-1} x^{2|s_a \mp m_a|+3/2};x^2)_\infty}{({z_a}^{\pm 1} {\alpha_a} x^{2|s_a \pm m_a|+1/2};x^2)_\infty} \end{align*}
Redefining parameters and introducing a new gauge group as before, this becomes:

\begin{align} \sum_{s,s_a \in \mathbb{Z}/2} \int \frac{dz}{2 \pi i z} \frac{dz_a}{2 \pi i z_a} z^{2n} w^{2 s} \prod_{a=1}^{N_f} (x^{1/2}  {\alpha_a}^{2})^{2|m_a|}  \frac{({\alpha_a}^{2} x^{4|m_a|+1};x^2)_\infty}{({\alpha_a}^{-2} x^{4|m_a|+1};x^2)_\infty} \times \end{align}

\begin{align*}\times  {z_a}^{2(n_a + s)} (z u_a)^{2s_a} (x^{1/2} {z_a}^{\pm 1} {\alpha_a}^{-1})^{|s_a \mp m_a|} \frac{({z_a}^{\pm 1} {\alpha_a}^{-1} x^{2|s_a \mp m_a|+3/2};x^2)_\infty}{({z_a}^{\pm 1} {\alpha_a} x^{2|s_a \pm m_a|+1/2};x^2)_\infty} \end{align*}

Recall that gauging this $U(1)_J$ symmetry should eliminate the gauge field.  In the case of the $S^3$ partition function, the effect was to introduce a delta function, and a similar thing happens here.  Namely, the integral over $z$ is only nonvanishing for $n + \sum_a s_a = 0$.  Similarly, the sum over $s$ gives:

\[ \sum_{s \in \mathbb{Z}/2} (w \prod_a z_a)^{2 s} \]
which is just a delta function which sets $w \prod_a z_a = 1$.  Thus the sum of all $N_f$ $U(1)$ gauge currents is ungauged, specifically, the sum of the gauge multiplets is set equal to the fixed value of the background vector multiplet correponding to the parameters $w$ and $n$.   

Finally, we need to redefine the gauge fields to get the standard presentation of the theory.  We define new variables $\hat{z}_a, \hat{s}_a$ by $z_a=\hat{z}_a{\hat{z}_{a+1}}^{-1} w^{-1/N_f}$ and $s_a=\hat{s}_a -\hat{s}_{a+1} - \frac{1}{N_f} n$.  Since there are fields with charge $\frac{1}{N_f}$ under the image of the $U(1)_J$ symmetry of SQED, $n$ must be a multiple of $N_f$ by gauge invariance, and so the $\hat{s}_a$ are integers.  Note that these new variables have a shift symmetry, which is most easily fixed by setting $\hat{z}_1=1$ and $\hat{s}_1=0$.  Then we get the index for the dual theory:

\begin{align} \sum_{\hat{s}_a \in \mathbb{Z}/2, \hat{s}_1=1} \int_{\hat{z}_1 = 1}  \frac{d\hat{z}_a}{2 \pi i \hat{z}_a} \prod_{a=1}^{N_f} (x^{1/2}  {\alpha_a}^{2})^{2|m_a|}  \frac{({\alpha_a}^{2} x^{4|m_a|+1};x^2)_\infty}{({\alpha_a}^{-2} x^{4|m_a|+1};x^2)_\infty} \times \end{align}

\[ \times  (\hat{z}_a{\hat{z}_{a+1}}^{-1} w^{-1/N_f})^{2 n_a} {u_a}^{2(\hat{s}_a -\hat{s}_{a+1} - \frac{1}{N_f} n)} (x^{1/2} (\hat{z}_a{\hat{z}_{a+1}}^{-1} w^{-1/N_f})^{\pm 1} {\alpha_a}^{-1})^{|\hat{s}_a -\hat{s}_{a+1} - \frac{1}{N_f} n \mp m_a|} \times \]

\[ \times \frac{((\hat{z}_a{\hat{z}_{a+1}}^{-1} w^{-1/N_f})^{\pm 1} {\alpha_a}^{-1} x^{2|\hat{s}_a -\hat{s}_{a+1} - \frac{1}{N_f} n \mp m_a|+3/2};x^2)_\infty}{({z_a}^{\pm 1} {\alpha_a} x^{2|\hat{s}_a -\hat{s}_{a+1} - \frac{1}{N_f} n \pm m_a|+1/2};x^2)_\infty} \]
It is straightforward to see that the symmetries map as expected.  Specifically, $(u_a,v_a)$, which parametrized the $U(1)_V$ symmetries of SQED, parameterize topological symmetries here, while $(w,n)$, which parametrized to $U(1)_J$ of SQED, corresponds to $-\frac{1}{N_f}$ times the overall $U(1)_V$.  The axial parameters $(\alpha_a,m_a$) map to axial parameters in the dual, up to a sign.

\appendix

\section{Verification of Leading Order Agreement at $m \neq 0$}

In this appendix we compare the generalized superconformal indices for $\mathcal{N}=2$ SQED and the XYZ for non-zero $m$ to lowest order in $x$.

The generalized superconformal index is given by:

\begin{align} I_{N_f=1}(\alpha,m;w,n;x) = \sum_{s \in \mathbb{Z}/2} \int \frac{dz}{2 \pi i z} z^{2n} w^{2s} (x^{1/2} z^{\pm 1} \alpha^{-1})^{|s \mp m|} \frac{(z^{\pm 1} \alpha^{-1} x^{2|s \mp m|+3/2};x^2)_\infty}{(z^{\pm 1} \alpha x^{2|s \pm m|+1/2};x^2)_\infty} \end{align}
As before, if we redefine parameters:

\begin{align} q=x^2, \;\;\; k = 2s, \;\;\; \ell = 2 n , \;\;\; a = \alpha^{-2} q^{1/2}, j = 2 m \end{align}
The index takes the form:

\begin{align} I_{N_f=1} = \sum_{k \in \mathbb{Z}} \int \frac{dz}{2 \pi i z} z^\ell w^k a^{(|k-j|+|k+j|)/4} z^{(|k-j|-|k+j|)/2}\frac{(z^{\pm 1} a^{1/2} q^{|k \mp j|/2+1/2};q)_\infty}{(z^{\pm 1} a^{-1/2} q^{|k \pm j|/2+1/2};q)_\infty} \end{align}

We wish to show this is equal to the generalized superconformal index of the XYZ theory, which, after identifying parameters, becomes:

\begin{align} I_{XYZ} = (a q^{-1/2})^{|j|} ( w^{\pm 1} a^{-1/2}q^{1/2})^{|j \pm \ell|/2} \frac{(a q^{|j|},w^{\pm 1} a^{-1/2} q^{|j \pm \ell|/2+1};q)_\infty}{(a^{-1} q^{|j|+1},w^{\pm 1} a^{1/2} q^{|j \mp \ell|/2};q)_\infty} \end{align}

We will give evidence for this equality by expanding both expressions in $q$.  First, it is convenient to define:

\begin{align} \epsilon=\sgn(\ell)\sgn(j) \end{align}

For the XYZ theory this expansion is straightforward, and we find:

\begin{align} I_{XYZ} = \left\{ \begin{array}{ccc} \displaystyle
a^{|j|/2} w^{\epsilon|\ell|} & & |\ell|<|j| \\& & \\ \displaystyle
\frac{w^{\epsilon|j|} a^{|j|/2}}{1 - w^{-\epsilon} a^{1/2}} & & |\ell|=|j| \\ & & \\ \displaystyle
a^{|j| - |\ell|/2} q^{(|\ell|-|j|)/2} w^{\epsilon |j|} & & |\ell|>j \end{array} \right. \end{align}

Now consider the expression for SQED:

\begin{align} \label{A7} \sum_{k \in \mathbb{Z}} \int \frac{dz}{2 \pi i z} z^\ell w^k a^{(|k-j|+|k+j|)/4} z^{(|k-j|-|k+j|)/2}\frac{(z^{\pm 1} a^{1/2} q^{|k \mp j|/2+1/2};q)_\infty}{(z^{\pm 1} a^{-1/2} q^{|k \pm j|/2+1/2};q)_\infty} \end{align}
We can keep only the first factor in the $q$-product, since the remaining ones introduce extra powers of $q$, and (\ref{A7}) becomes: 

\begin{align}\label{exp} \sum_{k \in \mathbb{Z}} \int \frac{dz}{2 \pi i z} z^{\ell + |k-j|/2-|k+j|/2} w^k a^{|k-j|/4+|k+j|/4} \frac{1 - z^{\pm 1} a^{1/2} q^{|k \mp j|/2+1/2}}{1 - z^{\pm 1} a^{-1/2} q^{|k \pm j|/2+1/2}} + ... \end{align}
The integration picks out the coefficient of $z^0$.  Let us define $\alpha=\ell + |k-j|/2-|k+j|/2$.  We need to consider three cases:

\begin{itemize}
\item $|\ell| < |j|$ - In this case, there is a unique value of $k$ such that $\alpha=0$.  Specifically, if we take $k=\sgn(j) \ell$, then $|k \pm j| = |\ell \pm |j|| = |j| \pm \ell$, and $\alpha$ vanishes.  When $\alpha=0$, the coefficient of $z^0$ is just:

\begin{align} w^k a^{|k-j|/4+|k+j|/4} \end{align}
Thus the expansion of the index in a power series in $q$ starts with a term of order $q^0$, and the leading contribution to the index is given by:

\begin{align} I_{N_f} =  w^{\epsilon |\ell|} a^{|j|/2} + ...\end{align}

\item $|\ell| = |j|$ - Now $\alpha=0$ whenever $|k| \geq |j|$ and $\sgn(k) = \sgn(j) \sgn(\ell)$.  Summing the resulting geometric series, we find the leading contribution is again of order $q^0$, and is given by:

\begin{align} I_{N_f} = \frac{w^{\epsilon |j|} a^{|j|/2}}{1 - w^{-\epsilon} a^{1/2}} + ... \end{align}

\item $|\ell| > |j|$ - Here it is no longer possible to make $\alpha=0$.  Now one must extract $-\alpha$ powers of $z$ from the following factor in (\ref{exp}):

\begin{align} \frac{1 - z^{\pm 1} a^{1/2} q^{|k \mp j|/2+1/2}}{1 - z^{\pm 1} a^{-1/2} q^{|k \pm j|/2+1/2}} \end{align}
in order to get a constant term. The $z^{-\alpha}$ term in the expansion of this fraction is given by, letting $\sgn(\alpha)=\pm$:

\begin{align}\label{term} ( z^{\mp 1} a^{-1/2} q^{|k \mp j|/2 +1/2})^{|\alpha|} ( 1 - a q^{|k \pm j|/2 - |k \mp j|/2}) \end{align} 

We wish to find the value of $k$ which minimizes the power of $q$ that appears.  Consider the exponent on $q$ that appears in the first term above:
\begin{align} (|k \mp j|/2 +1/2)(|\ell + |k-j|/2-|k+j|/2|) \end{align}
It is not hard to check that this function is linear for both $k>|j|$ and $k<-|j|$, and its slope is positive in the former case and negative in the latter, so the minimum must occur in the range $|k| \leq |j|$.  Using $\sgn(\alpha) = \sgn(\ell)$, we can rewrite this expression for $|k| \leq |j|$ as:
\begin{align} \frac{1}{2} (|j| + 1 - \epsilon k )(|\ell| - \epsilon k) \end{align}
This is quadratic, but the minimum occurs outside the region $|k| \leq |j|$, so the exponent is minimized at one of the boundary points, $k=\pm j$.  The correct choice is $k=\epsilon|j|$, giving an exponent of $(|\ell| - |j|)/2$.  Similarly, the second term in (\ref{term}) is minimized at the same $k$, and the exponent there is $(|\ell|+|j|)/2$.  Thus the former contribution is the leading one, and the index is given by:

\begin{align} I_{N_f} = w^{\epsilon |j|} a^{|j| - |\ell|/2} q^{|\ell|/2 - |j|/2} + ...\end{align}

\end{itemize}

Comparing the indices of the mirror theories for each case, we see they agree to leading order in $q$.

\acknowledgments{This work was supported in part by the DOE grant DE-FG02-92ER40701.}

\bibliographystyle{jhep}

\end{document}